\documentclass[aps,prd,twocolumn,twoside,preprintnumbers,nofootinbib,superscriptaddress,10pt]{revtex4-2}

\usepackage{amsmath}
\usepackage{graphicx}
\usepackage{dcolumn}
\usepackage{xspace}
\usepackage{color}
\usepackage{xcolor}
\usepackage{url}
\usepackage[utf8]{inputenc}
\usepackage{epstopdf}

\definecolor{nicered}{rgb}{0.8,0.1,0.1}
\definecolor{nicegreen}{rgb}{0.1,0.7,0.1}
\usepackage[colorlinks,citecolor=nicegreen ,linkcolor= nicered]{hyperref}
\usepackage{zref-clever}  
\zcsetup{cap=true,abbrev=true,nameinlink=false}
\zcRefTypeSetup{equation}{
  refbounds = {,(,),}
}
\zcLanguageSetup{english}{
  type = appendix,
  Name-sg = App.,  
  name-sg = app.,  
  Name-pl = Apps.,
  name-pl = apps.
}
\zcLanguageSetup{english}{
  type = section,
  Name-sg = Sec.,  
  name-sg = sec.,  
  Name-pl = Secs.,
  name-pl = secs.
}

\def\IFIC{Instituto de F\'isica Corpuscular,
Universitat de Val\`encia – Consejo Superior de Investigaciones Cient\'ificas,
Parc Cient\'ific, E-46980 Paterna, Valencia, Spain}

\def\UAB{Grup de F\'isica Te\`orica, Departament de F\'isica, Universitat Aut\`onoma de Barcelona, 
E-08193 Bellaterra (Barcelona), Spain}

\def\ICREA{ICREA, Instituci\'o Catalana de Recerca i Estudis Avan\c{c}ats,\\
Passeig de Llu\'{\i}s Companys 23, 
08010 Barcelona, Spain}

\def\IFAE{
Institut de F\'isica d'Altes Energies (IFAE), The Barcelona Institute of Science and Technology, 
Campus UAB, 08193 Bellaterra (Barcelona), Spain}

\begin{document}
\preprint{}

\title{
Impact of Hadronic Resonances on $B\to K^{(*)}\tau^+\tau^-$ Decays}

\author{Guillermo Balt\`a}
\email{gbalta@ifae.es}
\affiliation{\UAB}
\affiliation{\IFAE}

\author{Andreas Crivellin}
\email{andreas.crivellin@cern.ch}
\affiliation{\UAB}
\affiliation{\ICREA}

\author{Rafel Escribano}
\email{rescriba@ifae.es}
\affiliation{\UAB}
\affiliation{\IFAE}

\author{Joaquim Matias}
\email{matias@ifae.es}
\affiliation{\UAB}
\affiliation{\IFAE}

\author{Mart\'in Novoa-Brunet}
\email{martin.novoa@ific.uv.es}
\affiliation{\IFIC}

\begin{abstract}
Neutral-current semileptonic $B$ decays are plagued by hadronic resonances across the dilepton invariant-mass squared spectrum, $q^2$. For light leptons, $\ell=e,\mu$, these resonances can be avoided with suitable $q^2$ cuts. This strategy is less straightforward for $\tau$ modes, where missing energy from the $\tau$ decay makes $q^2$ difficult to reconstruct. In fact, while Belle II is able to discriminate between different regions in $q^2$ due to its clean environment, this is not directly possible in a hadronic one.
Therefore, the interpretation of $b\to s\tau^+\tau^-$ measurements from e.g.~LHCb, CMS requires the description of these resonant effects. In this article, we adopt a different strategy 
by including the resonant contributions (in particular from $\psi(2S)$) into our predictions for $B\to K^{(*)}\tau^+\tau^-$ decays, instead of avoiding them. We provide predictions for different initial kinematic points ($4m_\tau^2, 14.18$\,GeV$^2$ and $15$\,GeV$^2$) that can be convenient for LHCb, CMS and Belle II. For this, we use a data-driven approach based on the LHCb measurements of $B\to K^{(*)}\mu^+\mu^-$ decays. Including the resonances and integrating over the full $q^2$ range substantially enhances the Standard Model predictions. 
However, for sufficiently large New Physics, motivated by the current tensions in $R(D^{(*)})$ and $B\to K^{(*)}\nu\nu$ decays, the short-distance contribution becomes comparable to or even exceeds the resonant one. 
This highlights two advantages of this strategy: it exploits the additional phase space associated with the resonant regions to probe large New Physics contributions, and it enables the use of hadron-collider data, where the resonances cannot be resolved.
We further quantify how including or neglecting the resonances affects the total branching ratio as a function of New Physics contributions and, equivalently, of the experimental precision.
\end{abstract}

\maketitle

\section{Introduction}
\label{intro}

After the discovery of the Higgs boson~\cite{ATLAS:2012yve,CMS:2012qbp}, the search for physics beyond the Standard Model (SM) has become even more relevant. While no conclusive evidence for New Physics (NP) at particle colliders has been uncovered so far, interesting deviations from the SM predictions in semileptonic $B$-meson decays involving $b \to s\ell^+\ell^-$ and $b\to c \tau\nu$ transitions, the so-called $B$-flavour anomalies, have been observed (see Ref.~\cite{Capdevila:2023yhq}).

Concerning $b\to s\ell^+\ell^-$ transitions, the first anomaly was already found in 2013 in the angular observable $P_5^\prime$~\cite{Descotes-Genon:2012isb,Descotes-Genon:2013vna} in $B\to K^*\mu^+\mu^-$~\cite{LHCb:2013ghj,Descotes-Genon:2013wba}. Together with the differential $B\to K^{(*)}\mu^+\mu^-$ decays~\cite{
LHCb:2025mqb,LHCb:2014cxe}\footnote{In particular, the first lattice calculation over the full $q^2$ range from Ref.~\cite{Parrott:2022zte} leads to a tension of $>4\sigma$ in several bins of the $B^+\to K^+\mu^+\mu^-$ branching ratio (see Ref.~\cite{Alguero:2023jeh}).} and other angular observables in $B\to K^{*}\mu\mu$~\cite{
ATLAS:2018gqc,Belle:2016fev,LHCb:2014cxe,LHCb:2020gog,CMS:2024atz,LHCb:2025mqb}
and $B_s\to \phi\mu^+\mu^-$~\cite{LHCb:2021zwz,Gubernari:2022hxn,Gubernari:2020eft},
the global fit points towards NP with a significance above the $5\sigma$ level~\cite{Alguero:2023jeh,Hurth:2023jwr}. Interestingly, the tensions in $B_s\to \phi\mu^+\mu^-$ were recently confirmed by CMS~\cite{CMS:2026mhf}.

The anomalies in the ratios $R(D^{(*)})={\cal B}(B\to D^{(*)}\tau\nu)/{\cal B}(B\to D^{(*)}\ell\nu)$ ($\ell=e,\mu$) date back to 2012~\cite{BaBar:2012obs} and the measurements have been refined in the meantime~\cite{BaBar:2012obs,BaBar:2013mob,Belle:2015qfa,Belle:2016ure,Belle:2016dyj,Belle:2017ilt,Belle:2019rba,LHCb:2015gmp,LHCb:2017smo,LHCb:2017rln,LHCb:2023uiv,Belle-II:2025yjp,Belle-II:2026nmn,Belle-II:2026ICHEP}.\footnote{
At the Moriond 2026 conference in La Thuile and at ICHEP 2026, preliminary BaBar results using semileptonic $B$-tagging were presented which do not indicate an enhancement of these ratios~\cite{BaBar:2026MoriondQCD,ICHEPconference}; rather, they are consistent with the SM at the 1.5$\sigma$ level. However, since these results have not yet been published, we do not include them in our analysis. Nevertheless, it will be important to closely monitor future updates and the final publication of these measurements, as they may have a significant impact on the interpretation of the current experimental picture.} They point towards NP in the tau mode of the order of 10\% w.r.t.~the corresponding tree-level SM contribution with a significance of $\approx 3\sigma$ in case the two modes are considered to be independent~\cite{HeavyFlavorAveragingGroupHFLAV:2024ctg} and even $\approx 4\sigma$ in case one assumes the presence of NP in left-handed vector currents only~\cite{Crivellin:2025qsq}. 

Interestingly, there is an intrinsic correlation between these two anomalies in the SMEFT with manifest $SU(2)_L$ gauge invariance: the left-handed $b c\tau\nu$ current is linked to $b\to  s\tau^+\tau^-$ transitions. Furthermore, the latter generates a lepton-flavour universal effect in $C_9$ via a tau-loop effect through an off-shell photon penguin~\cite{Bobeth:2011st} with the right sign and magnitude preferred by $b\to s\ell^-\ell^+$~\cite{Crivellin:2018yvo}.

Importantly, in this setup, an enhancement of $b\to s\tau^+\tau^-$ processes of up to several orders of magnitude is predicted~\cite{Alonso:2015sja,Crivellin:2017zlb,Calibbi:2017qbu,Capdevila:2017iqn}. This expectation is consistent with current experimental limits~\cite{Belle:2021ecr,BaBar:2016wgb,LHCb:2025lcw,LHCb:2017myy,Belle-II:2025lwo,Belle:2026ooq}, with the most stringent limits at 90\% CL given by~\footnote{{The
$B\to K\tau^+\tau^-$ mode provides a less stringent bound than $B\to K^*\tau^+\tau^-$ since $K^*\to K^+\pi^-$ results in two charged tracks, allowing the reconstruction of the $B$  decay vertex.}}:
\begin{equation}
\begin{aligned}
{\cal{B}}{\left( {B_s \to {\tau ^ + }{\tau ^ - }} \right)_{{\rm{exp}}}} &\le 5.2\times 10^{-3}\;&&\text{LHCb~\cite{LHCb:2017myy}}
\,,\\
{\cal{B}}{\left( {B \to {K}{\tau^+\tau^-}} \right)_{{\rm{exp}}}} &\le 5.6 \times 10^{-4}\; &&\text{Belle + Belle\,II~\cite{Belle:2026ooq} }
\, \label{eq:explimits}  ,\\
{\cal{B}}{\left( {B \to {K^*}{\tau^+\tau^-}} \right)_{{\rm{exp}}}} &\le 2.5 \times 10^{-4}\;&&\text{LHCb~\cite{LHCb:2025lcw}}
\,.
\end{aligned}
\end{equation}

On the theory side, the SM prediction for the $B_s\to\tau^+\tau^-$ branching ratio is known precisely~\cite{Bobeth:2013uxa,Bobeth:2014tza} as the relevant hadronic input is limited to the well-controlled $B_s$ decay constant. Branching ratios for $B\to K^{(*)}\tau^+\tau^-$ were estimated to be of $\mathcal{O}(10^{-7})$~\cite{Hewett:1995dk,Bouchard:2013mia,Kamenik:2017ghi}. Ref.~\cite{Capdevila:2017iqn} calculated the rates for $B\to K^{(*)}\tau^+\tau^-$  within the $q^2$ window of 15\,GeV$^2$ up to the kinematic endpoint, such that hadronic resonances, in particular $\psi(2S)$, are excluded.\footnote{The corresponding $\Lambda_b\to \Lambda\tau^+\tau^-$ decay was studied in Ref.~\cite{Bordone:2025elp}, also excluding the resonance region.} 

However, the squared invariant mass of the lepton pair ($q^2$) cannot be measured directly for tau leptons due to the unavoidable missing energy in the final states in the form of neutrinos necessarily involved in their decays. Furthermore, while an indirect measurement via the $B$ momentum and the momentum of the final state meson is feasible at Belle II, where the other $B$ meson is tagged, this is not possible in a hadronic environment. Therefore, LHCb and CMS can only measure the $B\to K^{(*)}\tau^+\tau^-$ rates over the full $q^2$ range where hadronic resonances are present. However, no predictions for these measurements exist in the literature. 

In this article, we look at the $b\to s\tau^+\tau^-$ processes $B\to K^*\tau^+\tau^-$ and $B\to K\tau^+\tau^-$  to include the effects of these resonances via a data-driven approach based on recent LHCb measurements of the hadronic spectrum in the $B\to K^{(*)}\mu^+\mu^-$ modes~\cite{LHCb:2024onj,LHCb:2026suh,LHCb:2016due}. This is essential to provide the SM prediction for these modes in order to remain sensitive to smaller NP contributions.
Therefore, besides presenting the SM predictions, we also quantify the relative error of neglecting the resonant contributions as a function of NP. We provide expressions for the branching ratios in terms of the Wilson coefficients $C_{9\tau}^{\rm NP}$ and $C_{10\tau}^{\rm NP}$. 
Finally, we will illustrate our results for  ${\cal B}(B\to K^{(*)}\tau^+\tau^-)$ in a simple SMEFT setup by predicting their branching ratios as a function of $R(D^{(*)})$ and $B\to K^{(*)}\nu\nu$ in light of the respective anomalies.  

\section{Treatment of resonances}
\label{sec:resonances}

As discussed in the introduction, in a hadron-collider environment, such as at LHCb and CMS, 
it is not possible to reconstruct the invariant mass of the tau lepton pair. 
Therefore, a realistic prediction for the $B\to K^{(*)}\tau^+\tau^-$ branching ratio, 
which can be directly compared to the measurement, must include the full kinematically accessible $q^2$ range and, 
therefore, the resonance region.
Estimating these nonlocal hadronic effects, associated with intermediate charmonium states, is the main challenge to be addressed here.

For this, we adopt a data-driven approach based on the recent LHCb analyses of the decays $B^0\to K^{*0}\mu^+\mu^-$~\cite{LHCb:2024onj} 
and $B^+\to K^+\mu^+\mu^-$~\cite{LHCb:2026suh}, which provided the full $q^2$ spectrum. 
These analyses fit the local and nonlocal contributions, including charmonium resonances. 
In this way, the local short-distance contribution, the nonlocal hadronic contribution, 
and their interference are directly inferred from data. 
In addition, we assess the accuracy of using the narrow width approximation (NWA) 
and compare the obtained results to those of the full data-driven approach. 
We illustrate this comparison for both $B^+\to K^+\tau^+\tau^-$ and $B^0\to K^{*0}\tau^+\tau^-$ 
in Sec.~\ref{Results}.

\begin{figure*}  
\includegraphics[width=0.45\linewidth,height=6cm]{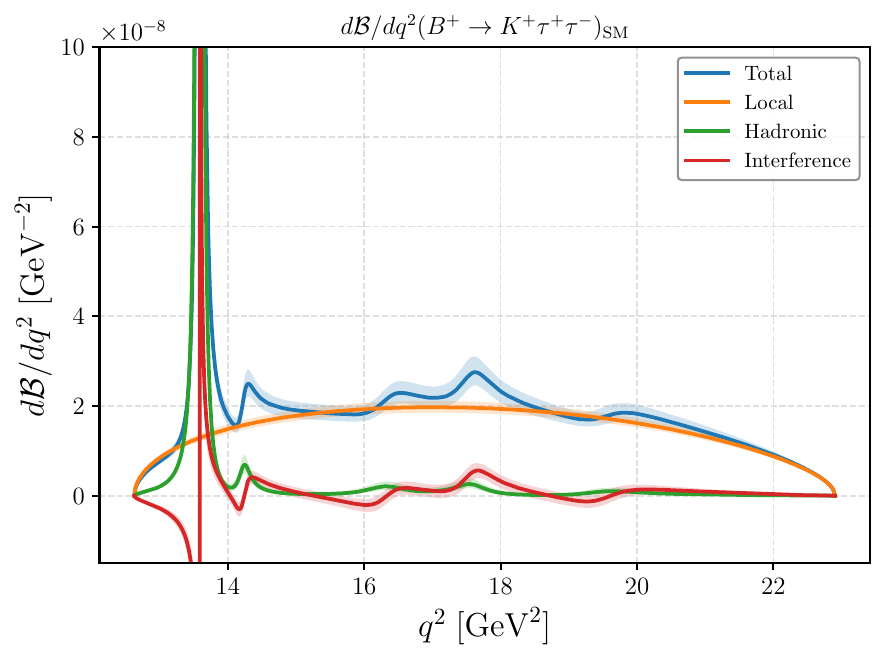}      
\includegraphics[width=0.45\linewidth]{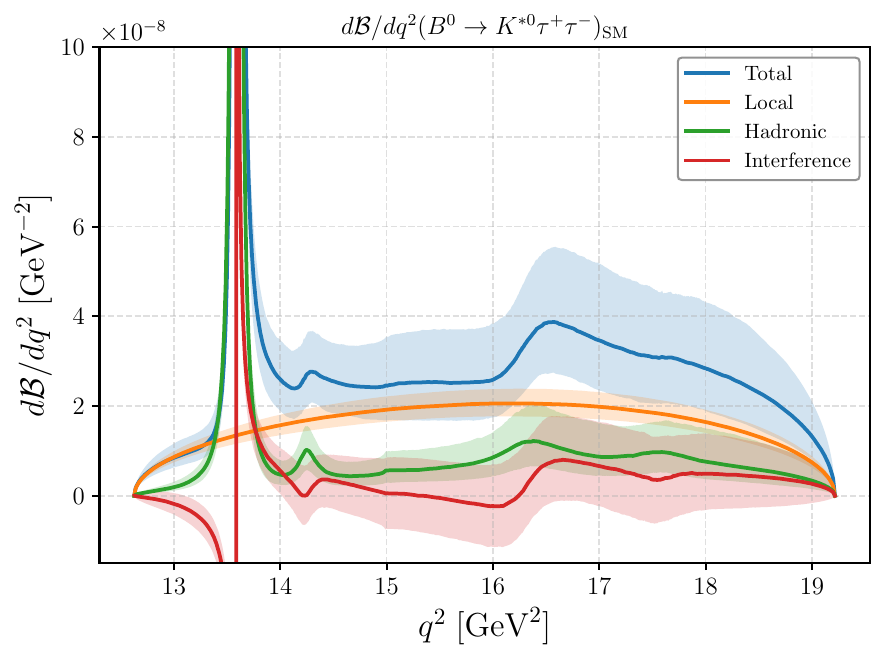}
\caption{Differential $B^+\to K^+\tau^+\tau^-$ and $B^0\to K^{*0}\tau^+\tau^-$ branching ratio central value and 
1$\sigma$ confidence interval within the SM (blue) decomposed into local contribution (orange), 
nonlocal hadronic effects (green) and their interference (red).}
\label{figshape}
\end{figure*}

The crucial point of our approach is that the nonlocal hadronic amplitudes are properties of the meson transitions 
(i.e.~$B\to K$ and $B\to K^*$) and, up to tiny QED effects, do not depend on the flavour of the final-state leptons. 
This allows us to transfer the nonlocal amplitudes (including their phases) extracted from $B\to K^{(*)}\mu^+\mu^-$ to the tauonic mode, 
while keeping the size of the relevant short-distance Wilson coefficients free, thereby allowing NP effects to be included.

\subsection{Narrow width approximation}
\label{subsec:psi2Sbenchmark}
Before turning to the data-driven approach, 
a first estimate of the nonlocal contribution can be obtained by considering the dominant $\psi(2S)$ resonance 
while neglecting other heavier $c\bar c$ resonances and additional nonlocal contributions. 
The width to mass ratio $\Gamma/m$ for this resonance is $\sim8\times10^{-5}$, 
meaning that it is sufficiently narrow to apply the NWA. In this approximation,
\begin{multline}
\label{eq:NWApsi2S}
\mathcal B(B\to K^{(*)}\tau^+\tau^-)|_{\rm nl}\simeq\\
\mathcal B(B\to K^{(*)}\psi(2S))\times \mathcal B(\psi(2S)\to\tau^+\tau^-)\ .
\end{multline}

In this context, note that the direct measurement of $\mathcal B(\psi(2S)\to\tau^+\tau^-)$~\cite{BES:2006jil} 
is known less precisely than the corresponding muonic one~\cite{ParticleDataGroup:2026mpi}.
Thus, the theory prediction can be improved by using the $\mathcal B(\psi(2S)\to \mu^+\mu^-)$ measurement~
and converting it to the tauonic case through the ratio of phase-space factors\footnote{A 
combination of both measurements could also be used, but the impact with current data is small.}
\begin{equation}
\frac{\mathcal B(\psi(2S)\to\tau^+\tau^-)}{\mathcal B(\psi(2S)\to\mu^+\mu^-)}
\simeq
\sqrt{1-\frac{4m_\tau^2}{m_\psi^2}}\left(1+\frac{2m_\tau^2}{m_{\psi}^2}\right).
\label{eq:Btautaufrommumu}
\end{equation}
Note that \zcref{eq:NWApsi2S} does not describe the lineshape of the resonance, but provides a good approximation of its size. 
However, to obtain a more realistic description of the resonant contributions, 
including their lineshapes and the interference among them, 
one can model this contribution using a data-driven approach based on a dispersion relation method and 
extract the magnitude and phase of each resonance from experimental data.

\subsection{Data-driven approach}
\label{subsec:datadriven}

Our procedure for the data-driven approach to include the hadronic effects is as follows:
\begin{enumerate}

\item We compute the local contributions in the standard way~\cite{Grinstein:1987vj,Buchalla:1995vs,Capdevila:2017bsm} 
following \zcref{app:BtoKRate,app:amplitudes}.
\item
We include the nonlocal hadronic amplitudes extracted from the fit to $B\to K^{(*)}\mu^+\mu^-$ data from 
Refs.~\cite{LHCb:2024onj,LHCb:2026suh} through a redefinition of $\mathcal{C}_{9\tau}$.

\begin{figure*}
\centering 
\includegraphics[width=0.45\linewidth]{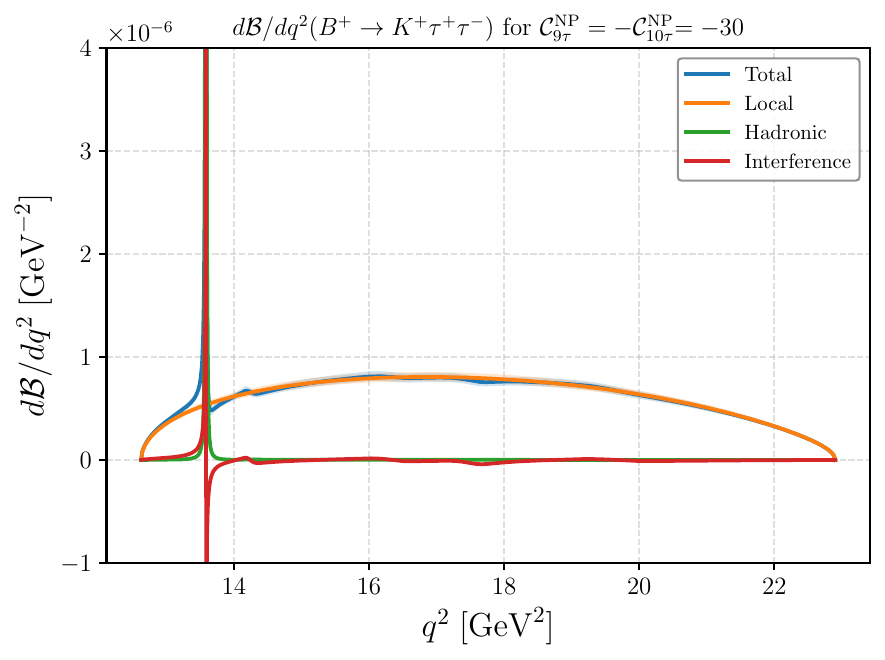}  
\includegraphics[width=0.45\linewidth]{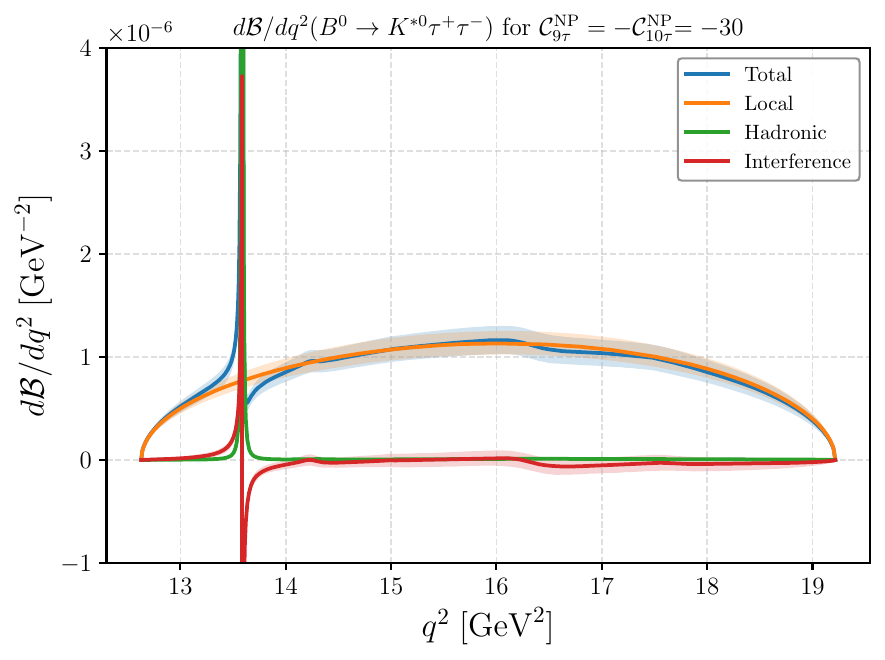}
\caption{
Same as \zcref{figshape} but for $\mathcal{C}_{9\tau}^{\rm NP}=-\mathcal{C}_{10\tau}^{\rm NP}=-30$.}
\label{figshapeNP}
\end{figure*}

\begin{figure*}
\centering
 \includegraphics[width=0.9\linewidth]{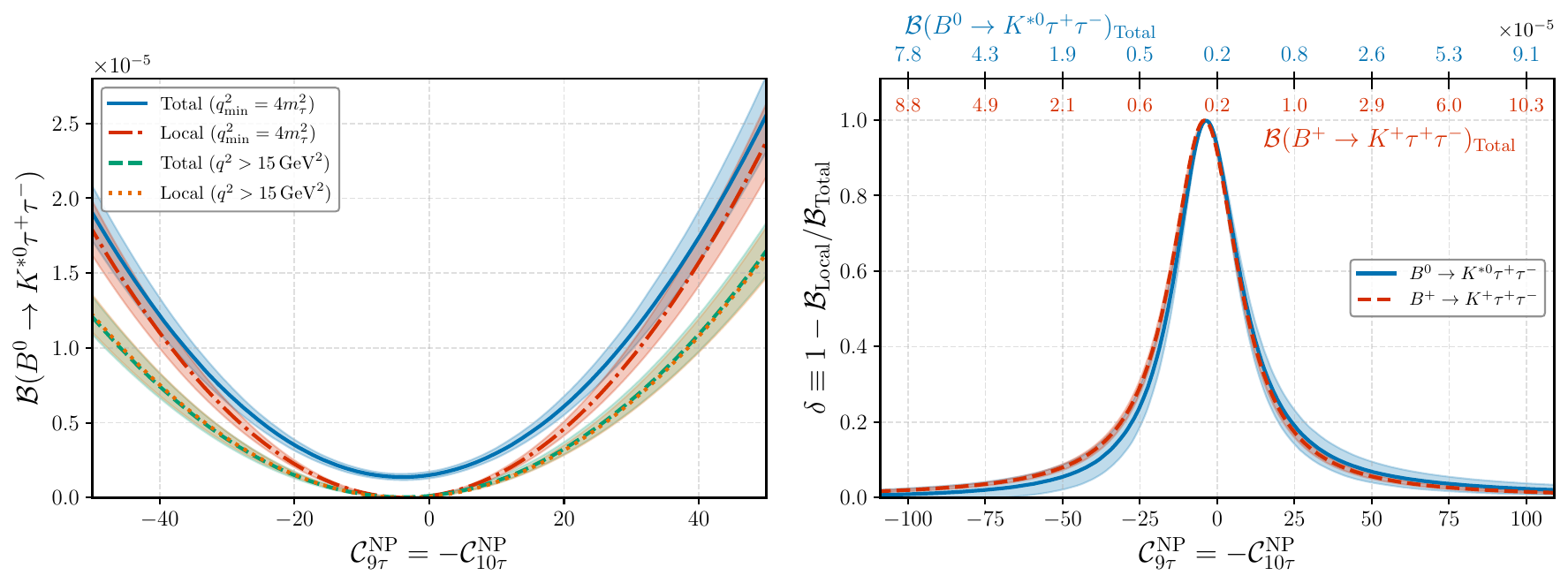}   
\caption{Left: Central value and $1\sigma$ interval for ${\cal B}(B^0\to K^{*0}\tau^+\tau^-)$ integrated from 
$4m_\tau^2$ ($15\,\mathrm{GeV}^2$) to the kinematic endpoint,
as a function of $\mathcal{C}_{9\tau}^{\rm NP}=-\mathcal{C}_{10\tau}^{\rm NP}$, including resonance effects in blue (green) 
and neglecting them in red (orange).
Right: Relative error $\delta$ made by neglecting resonance effects in the total branching fraction of $B^+\to K^+\tau^+\tau^-$ (red) 
and $B^0\to K^{*0}\tau^+\tau^-$ (blue), as a function of $\mathcal{C}_{9\tau}^{\rm NP}=-\mathcal{C}_{10\tau}^{\rm NP}$.
The colour-matched upper tick labels show the corresponding values of the branching fractions for $B\to K^{(*)}\tau^+\tau^-$
in units of $10^{-5}$.}
\label{fig:TotalandLocalvsNP}
\end{figure*}

\item
We integrate the spectra over the full kinematically allowed range
\begin{equation}
4m_\tau^2 \le q^2 \le (m_B-m_{K^{(*)}})^2\ ,
\end{equation}
or over bins starting at $14.18$ and $15$~GeV$^2$ extending up to the kinematic endpoint.
\end{enumerate}

This yields a prediction for the ditau mass spectrum and the corresponding branching ratio of the decays $B\to K^{(*)}\tau^+\tau^-$
that can be directly compared to the experimentally accessible data at LHCb and CMS.
In particular, these predictions include the effects of $c\bar c$ resonances\footnote{We 
include in the analysis the resonances $J/\psi$ (only the tail has an impact on the branching ratio by phase space), $\psi(2S)$, 
$\psi(3770)$, $\psi(4040)$, $\psi(4160)$, and $\psi(4415)$.
The light resonances $\rho(770)$, $\omega(782)$ and $\phi(1020)$ also considered in Ref.~\cite{LHCb:2026suh} for the $\mu^+\mu^-$ mode
are not within the kinematically allowed phase space for tau leptons.},
nonlocal hadronic effects due to $D\bar D^{(\ast)}$ loops considered in the LHCb analyses, 
and the interference of all these hadronic amplitudes with the short-distance amplitude.

\subsubsection{$B^+\to K^+\tau^+\tau^-$}

In the case of $B^+\to K^+\tau^+\tau^-$,
the ditau mass spectrum is described by a local short-distance contribution and a sum of nonlocal hadronic contributions,
each with a fitted magnitude and phase relative to the short-distance one.

Since the decay $B^+\to K^+\tau^+\tau^-$ involves a pseudoscalar final-state meson,
the nonlocal charm contributions enter via the effective Wilson coefficient 
\begin{equation}
\mathcal{C}_{9\tau}^{\rm eff}(q^2)=\mathcal{C}_{9\tau}+
Y_{c\bar c}^0(q_0^2)+\Delta Y_{c\bar c}^{\rm 1p}(q^2)+\Delta Y_{c\bar c}^{\rm 2p}(q^2)\ ,
\label{eq:C9effBK}
\end{equation}
where $\Delta Y_{c\bar c}^{\rm 1p}(q^2)$ and $\Delta Y_{c\bar c}^{\rm 2p}(q^2)$
are the one- and two-particle $c\bar c$ amplitudes obtained from once-subtracted dispersion relations,
with $Y_{c\bar c}^0(q_0^2)$ representing the subtraction constant evaluated at the subtraction point $q_0^2=0$.
The choice of this particular point and its calculation are explained in detail in Ref.~\cite{Bordone:2024hui}
and coincides with the point used in Ref.~\cite{LHCb:2026suh}.
The one-particle $c\bar c$ contribution is modelled as
\begin{equation}
\Delta Y_{c\bar c}^{\rm 1p}(q^2)=\sum_j\eta_j e^{i\delta_j}\frac{q^2-q_0^2}{m_j^2-q_0^2}A_j^{\rm res}(q^2)\ ,
\label{eq:res parametrization}
\end{equation}
where $\eta_j$ and $\delta_j$ are the magnitude and phase of the $j^{\rm th}$ resonance contribution
extracted from the fit to $B^+\to K^+\mu^+\mu^-$ data collected by the LHCb experiment~\cite{LHCb:2026suh}.
The factor $A_j^{\rm res}(q^2)$ is given by a relativistic Breit-Wigner (BW) function
\begin{equation}
A_j^{\rm res}(q^2)=
\frac{m_j\Gamma_{0j}}{m_j^2-q^2-i\,m_j\Gamma_j(q^2)}\ ,
\label{eq:Ajres}
\end{equation}
where $m_j$ and $\Gamma_{0j}$ represent the mass and natural width of each resonance, and $\Gamma_j(q^2)$ the running width.
To be consistent with the LHCb analysis, we keep the same energy dependence for the widths used there.
We also use the LHCb parametrization for the modelling of the two-particle nonlocal contributions.
For the local form factors, we rely on the results of the HPQCD collaboration~\cite{Parrott:2022rgu},
and use the $z$-parametrization described there.

The differential branching fraction is obtained from the standard expression given in \zcref{app:BtoKRate}
and is shown in \zcref{figshape}.
As seen, the contribution from the $\psi(2S)$ resonance is by far the dominant one, 
while the ones from the $J/\psi$ tail and the other heavier $c\bar c$ resonances are numerically negligible
(three orders of magnitude smaller).
The contribution from $D\bar D^{(\ast)}$ loops is of the same order and thus negligible as well.
In comparison with the $\psi(2S)$ resonant contribution,
the integrated local contribution in the $\tau^+\tau^-$ final-state case is only one order of magnitude smaller.
Due to the size of the different contributions besides the $\psi(2S)$ one,
the interference of the other $c\bar c$ resonances among themselves and with the short-distance contribution is negligible.
Only the interference of the $\psi(2S)$ and short-distance contributions is relevant in absolute terms.
However, since the contribution of the $\psi(2S)$ changes sign when passing through the $q^2=m_{\psi(2S)}^2$ value 
and the short-distance contribution is nearly constant around the narrow resonance region,
the net contribution of this interference is also negligible.
In summary, the main contributions to the integrated branching ratio of the $\tau^+\tau^-$ mode
come from the presence of the $\psi(2S)$ resonance itself and the short-distance ``background'',
rather than from other nonlocal contributions or interference effects.

\subsubsection{$B\to K^{{*}}\tau^+\tau^-$}

Here, the situation is more complicated compared to the case of $B\to K\ell^+\ell^-$, because resonance contributions depend on the $K^*$ helicity and therefore cannot be described by a universal shift $\mathcal{C}_{9\tau}\to \mathcal{C}_{9\tau}^{\rm eff}(q^2)$. Instead, they must be incorporated as helicity-dependent shifts at the level of the transversity amplitudes
\begin{equation}
\mathcal A_\lambda^{L,R}(q^2)
=
\mathcal A_{\lambda,\,{\rm loc}}^{L,R}(q^2)
+
\mathcal H_\lambda(q^2)\,,
\qquad
\lambda=\perp,\parallel,0\,,
\label{eq:Ahel}
\end{equation}
where $\mathcal A_{\lambda,\,{\rm loc}}^{L,R}$ denotes the local contribution given in \zcref{app:amplitudes} and $\mathcal H_\lambda(q^2)$ the nonlocal hadronic amplitude. 

The modelling of these nonlocal amplitudes, which follows the LHCb analysis of $B^0\to K^{*0}\mu^+\mu^-$~\cite{LHCb:2024onj}, is similar  to the $B^+\to K^+\tau^+\tau^-$ approach with two main differences. First, each one- or two-particle hadronic amplitude contributes differently to each of the three helicities. Second, no running widths are considered. We refer to Ref.~\cite{LHCb:2024onj} for further details.

For the local amplitudes, we use the form factors obtained from the dispersive analysis of Ref.~\cite{Gubernari:2023puw} instead of the posterior for the form-factor parameters obtained in Ref.~\cite{LHCb:2024onj}. We do so despite the fact that the form-factor parameters are, in principle, correlated with the extraction of the nonlocal contributions in Ref.~\cite{LHCb:2024onj}. We find, however, that the impact of these correlations is small. By contrast, using the corresponding form-factor posterior may induce a sizeable bias in the branching-fraction predictions in the presence of large NP contributions, if mismodelled nonlocal hadronic effects are absorbed into the form factors rather than into the nonlocal amplitude.

\subsection{Simplified data-driven approach in the NWA}
\label{subsec:simplified}

The model followed in the LHCb analysis for the $\mu^+\mu^-$ mode aims to describe in detail the shape of the differential branching ratio 
in the full kinematic range.
However, given the narrowness of the resonances involved and the current uncertainty of the theoretical predictions,
a simplified version of this model is possible and should be sufficient to provide a reliable prediction of the integrated branching ratio 
for the $\tau^+\tau^-$ mode, which could be useful as a cross-check of the LHCb results.

With this purpose in mind, we follow this alternative procedure,
which is based on the use of the same dispersive treatment of \zcref{eq:res parametrization}
with the same BW parametrization of \zcref{eq:Ajres},
but ignoring the energy dependence of the running widths $\Gamma_j(q^2)$ which are then fixed to the natural widths $\Gamma_{0j}$. 
The two-particle nonlocal contributions are not considered here because their impact is negligible with respect to 
the one-particle contributions.
For the resonances, 
the phases $\delta_j$ are still taken from the LHCb fit to data, 
while the magnitudes $\eta_j$ are calculated in the NWA, as shown in Ref.~\cite{Bordone:2024hui}.
In this way, the simplified model version is less dependent on the experimental data.
The results of this simplified approach are presented in the next section.

\section{Results}
\label{Results}

In the present section,
we present our predictions for the branching ratios $B\to K^{(*)} \tau^+\tau^-$,
including the nonlocal contributions using the data-driven approach based on the LHCb analyses.
Within the SM, the results over the full kinematic range are
\begin{equation}
\label{SM prediction}
\begin{array}{rcl}
{\cal B}(B^+\to K^+\tau^+\tau^-)_{\rm SM}&=& 1.86\substack{+0.17\\-0.16}\times 10^{-6}\ ,\\[1ex]
{\cal B}(B^0\to K^{*0}\tau^+\tau^-)_{\rm SM}&=& 1.50\substack{+0.23\\-0.22} \times 10^{-6}\ .
\end{array}
\end{equation}
In \zcref{tab:table1}, 
the corresponding SM predictions with different initial kinematic points $q^2_{\rm min}$  (in GeV$^2$) are provided.
These $q^2_{\rm min}$ are 15\,GeV$^2$,
which was used historically for the SM predictions~\cite{Capdevila:2017iqn} to veto the $\psi(2S)$ resonance,
and 14.18\,GeV$^2$, used by HPQCD~\cite{Parrott:2022zte} and Belle II in their experimental analysis~\cite{Belle:2026ooq}.
Furthermore, as shown in \zcref{figshape},
starting the integration at 15\,GeV$^2$ avoids the main contribution from the peak of the $\psi(2S)$ resonance,
resulting in a difference of one order of magnitude between the cases where the $\psi(2S)$ region is included or excluded.

\renewcommand{\arraystretch}{1.5}
\begin{table}
\centering
\begin{tabular}{|c|c|c|}
\hline $q^2_{\rm min}$ 
&
${\cal B}(B^+\to K^+\tau^+\tau^-)\times 10^7$ & ${\cal B}(B^0\to K^{*0}\tau^+\tau^-)\times 10^7$ \\
\hline 
14.18 & $1.54 \, \substack{+0.16 \\ -0.13}$ & $1.36 \, \substack{+0.61 \\ -0.40}$ \\
\hline
  15 & $1.37 \, \substack{+0.14 \\ -0.12}$ & $1.16 \, \substack{+0.54 \\ -0.36}$\\
\hline
\end{tabular}
\caption{SM predictions for the different initial kinematic points only accessible to Belle II.}
\label{tab:table1}
\end{table}

As a cross-check,
we  present the predictions for the branching ratios of $B^+\to K^+\tau^+\tau^-$ and $B^0\to K^{*0}\tau^+\tau^-$
using the approaches discussed in \zcref{subsec:psi2Sbenchmark,subsec:simplified} making use of \zcref{eq:Btautaufrommumu} with the muon mode as the normalization mode. Based on PDG averages~\cite{ParticleDataGroup:2026mpi} for the partial branching fractions the results are
\begin{itemize}
\item NWA (neglecting $\psi(2S)$-local interference effects):
\begin{equation}
\begin{array}{rcl}
{\cal B}(B^+\to K^+\tau^+\tau^-)&=& (2.10\pm 0.16)\times 10^{-6}\ ,\\[1ex]
{\cal B}(B^0\to K^{*0}\tau^+\tau^-)&=& (1.95\pm 0.20)\times 10^{-6}\ .
\end{array}
    \label{eq:BtoKtautauNWA} 
\end{equation}

\item Simplified data-driven approach using the NWA: 
 \begin{equation}
\label{eq:BtoKtautauSimplifiedDataDriven}
\begin{array}{rcl}
{\cal B}(B^+\to K^+\tau^+\tau^-)&=& (2.13\pm 0.16)\times 10^{-6}\ ,\\[1ex]
{\cal B}(B^0\to K^{*0}\tau^+\tau^-)&=& (1.99 \pm  0.20 )\times 10^{-6}\ .
\end{array}
\end{equation}
\end{itemize}

The good agreement between \zcref{eq:BtoKtautauNWA,eq:BtoKtautauSimplifiedDataDriven}
shows that the discrepancy between the simplified approach and the full data-driven approach
lies mainly in the extraction of the magnitudes of the resonances via NWA or from a fit to $B\to K^{(*)}\mu^+\mu^-$ data, as they rely on different datasets.
The interference between the local and nonlocal amplitudes, or the inclusion of the other $c\bar c$ resonances,
have, instead, a smaller effect. 

Furthermore, as can be seen from \zcref{figshapeNP}, even if large NP effects are turned on,
such that the magnitude of the interference
with the $\psi(2S)$ resonance becomes relevant,
the three approaches still agree well.
The reason is the cancellation between the constructive and destructive effects on both sides of the resonance region
when integrating over $q^2$, such that the net effect of the interference is negligible.

For observables integrated from $14.18\,{\rm GeV}^2$,
the situation is more delicate.
In this case, the cancellation between constructive and destructive interference around the resonance is not present,
and the result turns out to be
more sensitive to the relative phase between the local and nonlocal amplitudes.
If this phase is left unconstrained, the corresponding uncertainty can reach about $30\%$.
Therefore, once experimental sensitivities approach branching fractions of order $10^{-7}$,
interference effects in this region should be treated with particular care.
Similar effects have been pointed out in related baryonic modes~\cite{Bordone:2025elp}.

We now look at the numerical effect of including the $\psi(2S)$ resonance,
shown in \zcref{fig:TotalandLocalvsNP} (left),
in the prediction for $B^0\to K^{*0}\tau^+\tau^-$ in the possible presence of NP effects.
For this, we consider three different cases:
(i) integration over the full phase space including the resonance (Total, blue);
(ii) integration over the full phase space disregarding the resonance (Local, red);
(iii) cutting at 15\,GeV$^2$ to remove the resonance effect.
As one can see from \zcref{fig:TotalandLocalvsNP},
if NP is small, i.e.~similar to the short-distance contribution of the SM, including the resonance, then
(i) increases the branching ratio by an order of magnitude, compared to both (ii) and (iii).
If NP significantly exceeds the short-distance SM contribution by roughly a factor of 10, 
then the resonance effect can be neglected, and (i) and (ii) converge.
However, (i) and (iii) always disagree, and the difference increases with the size of NP.

To quantify the error induced by disregarding the resonance contribution,
we define
\begin{equation}
\delta\equiv 1-\frac{{\cal B}_{\rm Local}}{{\cal B}_{\rm Total}}\ ,
\end{equation}
which we show in \zcref{fig:TotalandLocalvsNP} (right),
as a function of the NP contribution ($\mathcal{C}_{9\tau}^{\rm NP}=-\mathcal{C}_{10\tau}^{\rm NP}$)
and equivalently of its effect on the branching fraction (upper axis).

Here, one can see that in the SM case, $\delta$ is around 90\%.
For moderate NP of about five times the local SM contribution,
the error remains around 20\%, while for very large NP (see \zcref{sec:SMEFT}) the resonance effect can be neglected. 

Finally,
we give semi-analytic expressions for the branching ratios as a function of the NP Wilson coefficients ${\cal C}^{\rm NP}_{9\tau,10\tau}$
in \zcref{app:semi_analytical}.

\subsection{SMEFT and correlations with $R(D^{(*)})$ and $B\to K^{(*)}\nu\nu$}
\label{sec:SMEFT}

\begin{figure}[t]
    \centering
    \includegraphics[width=\linewidth]{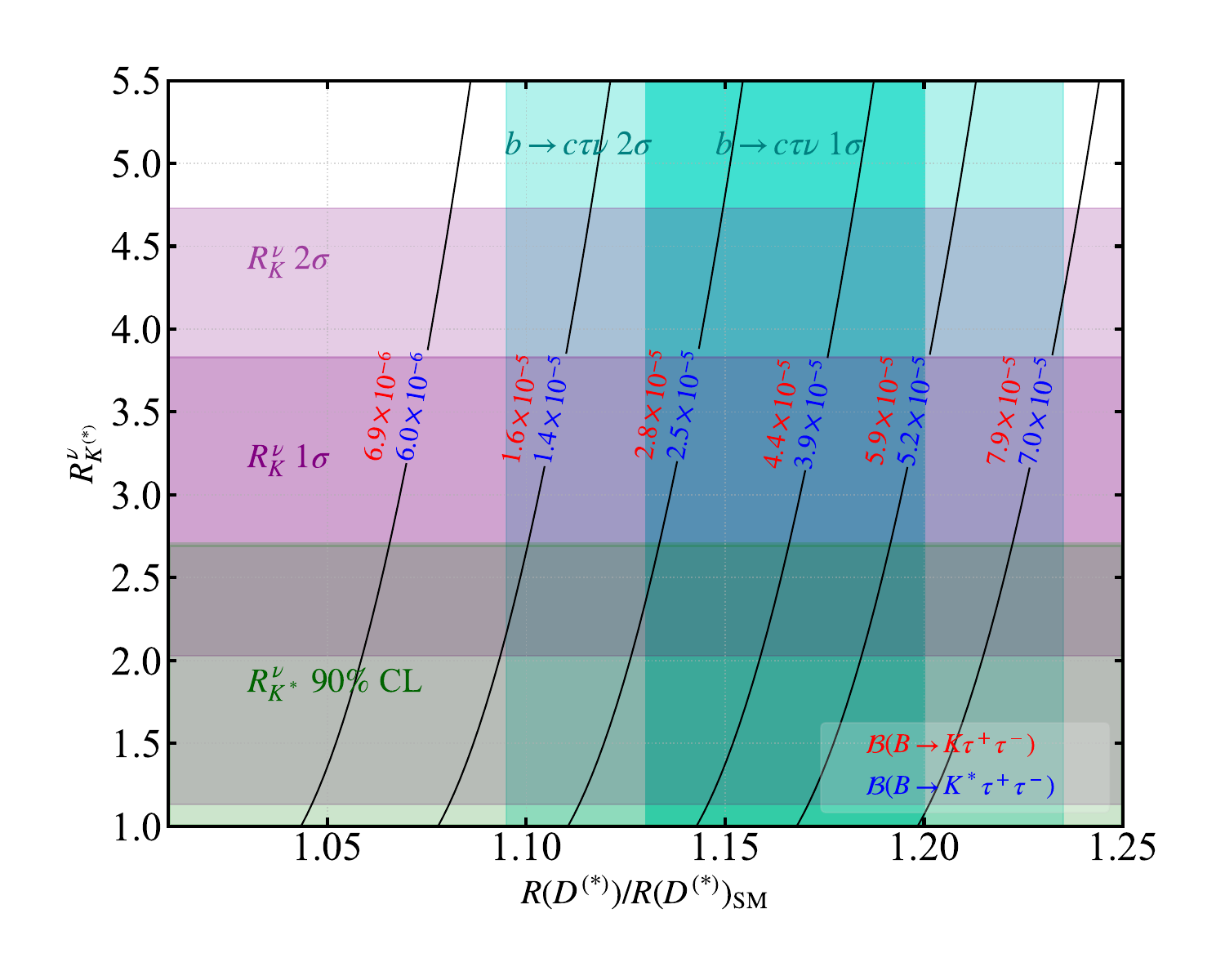}
    \caption{Predictions for ${\cal B}(B\to K^{(*)}\tau^+\tau^-)$ as a function of $R(D^{(*)})/R(D^{(*)})_{\rm SM}$ and $R_{K^{(*)}}^\nu$ in the scenario with only left-handed vector operators with a generic (non-hierarchical) flavour structure. Vertical light blue bands are preferred by $b\to c\tau\nu$ measurements, horizontal purple bands indicate the region preferred by $R_K^\nu$~\cite{Belle-II:2023esi}, while the green region is consistent with $R_{K^*}^\nu$ at $90\%$ CL~\cite{Belle:2017oht}\footnote{$R_K^\nu$ and $R_{K^*}^\nu$ have been calculated using the SM predictions from \cite{Becirevic:2023aov} and \cite{Buras:2014fpa}, respectively. The former value does not include the tree-level contribution~\cite{Kamenik:2009kc} which is subtracted in the experimental analysis~\cite{Belle-II:2023esi}.}. Note that while there are in principle four solutions for the branching ratio in terms of $R(D^{(*)})/R(D^{(*)})_{\rm SM}$ and $R_{K^{(*)}}^\nu$, we only show the one where no overcompensation of the SM occurs.}
    \label{fig:BRsRD_RKnuSol1}
\end{figure}

Finally, let us illustrate our results for $B\to K^{(*)}\tau^+\tau^-$ in the SMEFT~\cite{Buchmuller:1985jz} where direct correlations with $R(D^{(*)})$ and $B\to K^{(*)}\nu\nu$ exist in a simple setup with left-handed currents and generic (non-hierarchical) Wilson coefficients. 

To be more specific, the two operators 
\begin{equation}
\begin{array}{rcl}
{\cal O}^{(1)}_{2333}&=&[{\bar Q}_2 \gamma_\mu Q_3] [{\bar L}_3 \gamma^\mu L_3]\ ,  \\[1ex]
{\cal O}^{(3)}_{2333}&=&[{\bar Q}_2 \gamma_\mu \sigma^I Q_3] [{\bar L}_3 \gamma^\mu \sigma^I L_3]\ ,
\end{array} 
\end{equation}
are known to be capable of explaining $R(D^{(*)})$ without violating the bounds from EW precision data, LHC constraints and the $B_c$ lifetime\footnote{Indirect constraints on $b\to s\tau^+\tau^-$ operators are in general loose once the effects in $b \to s\tau^+\tau^-$ and $b \to d\tau^+\tau^-$ transitions are correlated such that the stringent bounds from $\Delta \Gamma_s/\Delta \Gamma_d$ are avoided~\cite{Bobeth:2014rda}. These operators are generated most naturally either in the $U_1$ vector-leptoquark model~\cite{Calibbi:2015kma,Barbieri:2016las,DiLuzio:2017vat,Calibbi:2017qbu,Bordone:2017bld,Blanke:2018sro,Crivellin:2018yvo} or in the $S_1+S_3$ scalar leptoquark model~\cite{Crivellin:2017zlb,Crivellin:2019dwb,Gherardi:2020qhc}. Note that while the former predicts $C^{(1)}=C^{(3)}$ at tree-level, ensuring the smallness of the $b\to s\nu\nu$ effect, the latter needs some tuning to keep the respective contributions at a phenomenologically acceptable level.}. Furthermore, they lead to the tau-loop induced effect in ${\cal C}_9^{\rm U}$, i.e.~a LFU effect in $b\to s\ell^+\ell^-$, discussed in the introduction.

Note that $C^{(3)}_{2333}$ contributes to $b\to c\tau\nu$ processes while $B\to K^{(*)}\nu\nu$ is sensitive to $C^{(1)}_{2333}-C^{(3)}_{2333}$. Furthermore, since only left-handed vector currents are present, all $b\to c\tau\nu$ and $b\to s\nu\nu$ processes are rescaled in the same way, such that 
\begin{equation}
  \frac{R(D^{*})}{R(D^{*})_{\rm SM}}=\frac{R(D)}{R(D)_{\rm SM}}
  \,,\quad R_{K^{*}}^\nu=R_K^\nu\,,  
\end{equation}
where
\begin{equation}
 R_{K^{(*)}}^{\nu}\equiv \frac{{\cal B}(B\to K^{(*)}\nu\nu)}{{\cal B}(B\to K^{(*)}\nu\nu)_{\rm SM}}\,.
\end{equation}
Therefore, if we neglect small CKM rotations (i.e.~$C^{(1,3)}_{3333} V_{cb}\ll C^{(1,3)}_{2333}$) we can express any $b\to s\tau^+\tau^-$ process in terms of these two ratios~\cite{Balta:2026fnz}.

This is illustrated in \zcref{fig:BRsRD_RKnuSol1} where ${\cal B}(B\to K^{(*)}\tau^+\tau^-)$ is shown as a function of $R(D^{(\ast)})/R(D^{(\ast)})_{\rm SM}$ and $R_{K^{(*)}}^{\nu}$. Regarding the importance of the resonant contribution, even in the lower left part of the figure, where NP effects are small, the local (enhanced by NP) and nonlocal contributions are already comparable. Moving towards the middle and right parts of the figure, the local part is large enough such that the nonlocal part becomes negligible. We can see that for the currently preferred values of $R(D^{(\ast)})$~\cite{HeavyFlavorAveragingGroupHFLAV:2024ctg} and $R_K^{\nu}$
~\cite{Belle-II:2023esi,Belle:2019iji,Belle-II:2021rof,Buras:2014fpa}, ${\cal B}(B\to K^{(*)}\tau^+\tau^-)$ is predicted to be a few times $10^{-5}$. In that region, $\delta\lesssim0.05$, thus ${\cal B}_{\rm Local}\simeq{\cal B}_{\rm Total}$.

\section{Conclusions and Outlook}

In this article, we have examined the effects of hadronic resonances on the theory predictions for $B\to K^{(*)}\tau^+\tau^-$ within and beyond the SM. This is necessary since hadron-collider experiments like LHCb and CMS cannot reconstruct the invariant mass of the tau-lepton pair with sufficient precision, such that (unlike the case of light leptons) one cannot constrain the phase space to exclude the resonances. 

We address the problem of including the resonances with a data-driven approach using the LHCb analyses of $B\to K^*\mu^+\mu^-$~\cite{LHCb:2024onj} and $B^+\to K^+\mu^+\mu^-$~\cite{LHCb:2026suh} decays, which provide fits to the spectra over the full $q^2$ region. The only numerically relevant resonance in the kinematically accessible $q^2$ range is $\psi(2S)$, which is narrow. In fact, we show that a Breit-Wigner parametrization and even a narrow-width approximation recover the full data-driven result for the integrated branching ratio well. 

For the SM, including the effect of $\psi(2S)$ leads to an enhancement of the branching ratio by an order of magnitude:
\begin{equation}
\begin{array}{rcl}
{\cal B}(B^+\to K^+\tau^+\tau^-)_{\rm SM}&=& 1.86\substack{+0.17\\-0.16}\times 10^{-6}\ ,\nonumber\\[1ex]
{\cal B}(B\to K^*\tau^+\tau^-)_{\rm SM}&=& 1.50\substack{+0.23\\-0.22} \times 10^{-6}\ ,
\end{array}
\end{equation}
while for large NP contributions, the resonant effect becomes small compared to the local contribution. However, even if NP is dominant over the (short and long distance) SM contribution (including the $\psi(2S)$ resonance), integrating over the full phase space instead of cutting on $q^2$ to remove the $\psi(2S)$ resonance has a significant numerical impact. To be more specific, even when NP fully dominates the decay $B\to K^{(*)}\tau^+\tau^-$, integrating over $q^2>4m_\tau^2$ yields a branching ratio around 50\% larger than that obtained with $q^2>15\,$GeV$^2$ simply because of the larger available phase space. Therefore, it is important to use the branching ratios defined over the full phase space for comparison with the measurements of LHCb and CMS.

In a SMEFT approach, we illustrate the predictions for $B\to K^{(*)}\tau^+\tau^-$ as a function of $R(D^{(*)})$ and $B\to K^{(*)}\nu\nu$, where interesting anomalies have been observed which point towards a sizable enhancement of $b\to s\tau^+\tau^-$ processes. Finally, note that our approach can also be applied to $B_s\to \phi\tau^+\tau^-$ and $\Lambda_b\to \Lambda\tau^+\tau^-$~\cite{Bordone:2025elp}, where again the $\psi(2S)$ resonance has a relevant impact. In fact, these two modes cannot be measured at Belle II, such that predictions over the full $q^2$ range are essential to compare them directly to the measurements.

\acknowledgments{We thank Kostas Petridis, Paula Alvarez-Cartelle, Mitesh Patel and Hanae Tilquin
for very useful discussions. 
We also thank Bernat Capdevila for contributing to the numerical implementation.
J.M.~gratefully acknowledges the financial support from ICREA under the ICREA Academia programme 2018
and to AGAUR under the Icrea Academia programme 2024, and from the IPPP Diva Award 2024. 
G.B., R.E.~and J.M.~received financial support from the
Spanish Ministry of Science, Innovation and Universities (project PID2023-146142NB-I00)
and Grant CEX2024-001441-S funded by MICIU/AEI/10.13039/501100011033.
M.N.~acknowledges financial support of the Spanish Government
(Agencia Estatal de Investigaci\'on MCIN/AEI/10.13039/501100011033), ERDF/EU
and the European Union NextGenerationEU/PRTR through the ``Juan de la
Cierva'' programme (Grant No. JDC2022-048787-I), and Grants
No.~PID2023-146220NB-I00 and No.~CEX2023-001292-S.
R.E.~and J.M.~are Serra Húnter Fellows.}

\appendix

\section{$B\to K\ell^+\ell^-$ rate}
\label{app:BtoKRate}
Differential branching ratio for $B^+\to K^+\tau^+\tau^-$

\begin{widetext}
\begin{equation}
\begin{split}
\frac{d\mathcal B(B^+\to K^+\tau^+\tau^-)}{dq^2}&=
\tau_B
\frac{G_F^2\alpha^2}{1024\pi^5m_B^3}
|V_{tb}V_{ts}^*|^2
\sqrt{\lambda(q^2)}\,\beta_\tau(q^2)
\Bigg[
\frac{2}{3}\lambda(q^2)\beta_\tau^2(q^2)
\left|(\mathcal{C}_{10\tau}+\mathcal{C}_{10'\tau})f_+(q^2)\right|^2
\\
&\hspace{0.6cm}
+\frac{4 m_\tau^2(m_B^2-m_K^2)^2}{q^2}
\left|(\mathcal{C}_{10\tau}+\mathcal{C}_{10'\tau})f_0(q^2)\right|^2
\\
&\hspace{0.6cm}
+\lambda(q^2)\left(1-\frac{1}{3}\beta_\tau^2(q^2)\right)
\left|
(\mathcal{C}_{9\tau}^{\rm eff}(q^2)+\mathcal{C}_{9'\tau})f_+(q^2)
+2(\mathcal{C}_7^{\rm eff}+\mathcal{C}_{7'}^{\rm eff})\frac{m_b+m_s}{m_B+m_K}f_T(q^2)
\right|^2
\Bigg] ,
\end{split}
\label{eq:dBrBKtautau}
\end{equation}
\end{widetext}
where $\lambda(q^2)=
\left(m_B^2+m_K^2-q^2\right)^2-4m_B^2m_K^2$ is the K\"{a}ll\'en function and $\beta_\tau(q^2) = \sqrt{1 - 4m_\tau^2/q^2}$. The form factors $f_{+,0,T}(q^2)$ are taken from Ref.~\cite{Parrott:2022rgu}.

\section{$B\to K^*\ell^+\ell^-$ Amplitudes}
\label{app:amplitudes}

For the local amplitudes, we use the convention
\begin{equation}
\mathcal A_{\lambda,\,{\rm loc}}^{L,R}(q^2)
=
\mathcal{N}_\lambda\left[
\left(\mathcal{C}_{9\lambda}^{\rm eff}(q^2)\mp \mathcal{C}_{10\lambda}\right)
F_\lambda^V(q^2)
+
\mathcal{C}_{7\lambda}^{\rm eff}F_\lambda^T(q^2)
\right],
\label{eq:Aloc}
\end{equation}
with $\lambda=\perp,\parallel,0$ and
\begin{equation}
C_{i\lambda}^{\rm (eff)}
=
C_i^{\rm (eff)}
+
\sigma_\lambda C_i^{\prime}\,,
\quad
\sigma_\lambda=\{1,-1,-1\}\ ,
\label{eq:Cilambda}
\end{equation}
for $i=7,9,10$, while the corresponding the transversity-dependent normalization factors are
\begin{equation}
    \mathcal{N}_\lambda=N\,\left\{\sqrt{2\lambda(q^2)},-\sqrt{2}(m_B^2-m_{K^*}^2),\frac{-1}{2 m_{K^*}\sqrt{q^2}}\right\}\ .
\end{equation}
Here $N = V_{tb} V_{ts}^* \sqrt{\frac{G_F^2 \alpha_{\rm em}^2 q^2 \lambda^{1/2}(q^2) \beta_\tau(q^2)}{3072 \pi^5 m_B^3}}$ and the Källén function is $\lambda(q^2)=
\left(m_B^2+m_{K^*}^2-q^2\right)^2-4m_B^2m_{K^*}^2$.

The transversity-dependent kinematic factors are absorbed into the hadronic functions
$F_\lambda^V$ and $F_\lambda^T$, which we define as
\begin{align}
\!\!\!\!\!F_\lambda^V(q^2) &\equiv \left\{\frac{V(q^2)}{m_B+m_{K^*}}, \frac{A_1(q^2)}{m_B-m_{K^*}}, 
\frac{A_{12}(q^2)}{m_B-m_{K^*}}\right\}\ ,
\end{align}
and
\begin{align}
F_\lambda^T(q^2) &\equiv  \frac{2m_b}{q^2}\left\{T_1(q^2), T_2(q^2), T_{23}(q^2)\right\}\ ,
\end{align}
where $A_{12}$ and $T_{23}$ are defined\footnote{These definitions differ from Refs.~\cite{LHCb:2024onj,Gubernari:2023puw} by kinematic factors.}
in Ref.~\cite{Descotes-Genon:2013vna}.

\section{Semi-analytical expressions}
\label{app:semi_analytical}
The explicit expressions of the total $B\to K^{(*)}\tau^+\tau^-$ branching ratios as a function of the NP Wilson coefficients $C_{9\tau,10\tau}^{\rm NP}$ are
\begin{widetext}

\begin{equation}\label{BtoK vs NP}
\mathcal{B}(B^+\to K^+\tau^+\tau^-)_{\text{Total}} =
1.85 \times 10^{-6}
+ (2.1  \mathcal{C}_{9\tau}^{\rm NP}
-5.4 \mathcal{C}_{10\tau}^{\rm NP}
+ 0.31   {\mathcal{C}_{9\tau}^{\rm NP}}^2
+ 0.62 {\mathcal{C}_{10\tau}^{\rm NP}}^2)\times 10^{-8}\, ,
\end{equation}

\begin{equation}
\label{BtoKstar vs NP}
\mathcal{B}(B^0\to K^{*0}\tau^+\tau^-)_{\text{Total}} =
1.48 \times 10^{-6}
+ (4.9 \mathcal{C}_{9\tau}^{\rm NP}
- 1.7 \mathcal{C}_{10\tau}^{\rm NP}
+ 0.63 {\mathcal{C}_{9\tau}^{\rm NP}}^2
+ 0.20 {\mathcal{C}_{10\tau}^{\rm NP}}^2
)\times 10^{-8}\,.
\end{equation}
Notice that the small discrepancies between the SM predictions of~\zcref{SM prediction} 
and~\zcref{BtoK vs NP,BtoKstar vs NP} 
are due to the fact that the former are derived from the median of the distribution, 
while the latter use central values of the input parameters.
    
\end{widetext}

\bibliography{bstautau_biblio}
\end{document}